\documentclass[aps,graphicx,twocolumn]{revtex4}
\usepackage{graphicx}

\begin{document}

\title{Deterministic entanglement purification and complete nonlocal Bell-state analysis with hyperentanglement\footnote{Published
in Phys. Rev. A \textbf{81}, 032307 (2010)}}

\author{Yu-Bo Sheng$^{1,2,3}$  and Fu-Guo Deng$^{1\footnote{Corresponding author:fgdeng@bnu.edu.cn.}}$}
\address{$^1$ Department of Physics, Beijing Normal University, Beijing 100875, People's Republic of China\\
$^2$ College of Nuclear Science and Technology, Beijing Normal
University, Beijing 100875, People's Republic of China\\
$^3$ Key Laboratory of Beam Technology and Material Modification of
Ministry of Education, Beijing Normal University, Beijing 100875,
People's Republic of China}
\date{\today }

\begin{abstract}
Entanglement purification is a very important element for
long-distance quantum communication. Different from all the existing
entanglement purification protocols (EPPs) in which   two parties
can only obtain some  quantum systems in a mixed entangled state
 with a higher fidelity probabilistically by consuming quantum
resources exponentially, here we present a deterministic EPP with
hyperentanglement. Using this protocl, the two parties can, in
principle, obtain deterministically maximally entangled pure states
in polarization without destroying any less-entangled photon pair,
which will improve the efficiency of long-distance quantum
communication exponentially. Meanwhile, it will be shown that this
EPP can be used to complete nonlocal Bell-state analysis perfectly.
We also discuss this EPP in a practical transmission.
\end{abstract}
 \maketitle

\section{introduction}

The realization of long-distance quantum communication schemes
should resort to the distribution of entangled states between
distant locations \cite{rmp}. Although photons are the optimal
quantum information carriers in long-distance quantum communication
as the interaction between them and environment is weaker than
others, the polarization degree of freedom of photons is incident to
 the noise in a quantum channel. Noise
will degrade the entanglement of a photon pair or even turn it into
a mixed state. If the destructive effect of the noise is low, the
two parties in quantum communication, say Alice and Bob, can first
exploit entanglement purification to improve the entanglement of the
quantum systems, and then achieve the goal of quantum communication
with maximally entangled states. Entanglement purification becomes a
very important element in quantum repeater \cite{repeater} for
long-distance quantum communication.

In 1996,  Bennett \emph{et al.} \cite{Bennett1} proposed an
entanglement purification protocol (EPP) based on quantum
controlled-NOT (CNOT) logic operations, and subsequently it was
improved by Deutsch \emph{et al.} \cite{Deutsch} using similar logic
operations. In 2001, Pan \emph{et al.} \cite{Pan1} proposed an EPP
with linear optical elements. In 2002, Simon and Pan \cite{Simon}
improved their protocol. They considered a currently available
source, a parametric down-conversion (PDC) source, to prepare
entangled photon pairs, and they first used spatial entanglement to
purify polarization entanglement. Both of these protocols should
resort to sophisticated single-photon detectors, which is not a
simple task in linear optics. In 2008, an EPP based on
nondestructive quantum nondemolition detectors was proposed
\cite{shengpra}. By far, all existing EPPs cannot obtain maximally
entangled states. They only improve the fidelity of an ensemble in a
mixed entangled state. In order to obtain some entangled states with
higher fidelity, they have to consume more and more
less-entanglement ones. Theoretically speaking, it is impossible to
get a pair of photons in a maximally entangled  state with
conventional EPPs \cite{Bennett1,Deutsch,Pan1,Simon,shengpra}.

Recently, the applications of  hyperentangled states have been
studied by some groups.  A state of being simultaneously entangled
in multiple degrees of freedom is  called "a hyperentangled state"
\cite{hyper-generation1,hyper-generation2,hyper3}. The most
important use of hyperentanglements is in a complete deterministic
local Bell-state analysis
\cite{hyper-analysis2,hyper-analysis3,hyper-analysis4}. In 2008,
with the help of the hyperentangled state in both polarization and
orbit angular momentum, Barreiro \emph{et al.} \cite{hyper6} beat
the channel capacity limit for linear photonic superdense coding.
With a type-I and type-II $\beta$ barium borate (BBO) crystal,
photon pairs produced by spontaneous parametric down-conversion
(SPDC) can be in the hyperentangled state in polarization and
spatial degrees of freedom \cite{Simon} polarization, spatial,
energy, and  time degrees of freedom \cite{hyper-generation1},
polarization and frequency degrees of freedom
\cite{Hyper-fruquency}; and so on.
 In 2009, Vallone \emph{et. al.}
\cite{hyperentangled2} also reported their experiment with a
six-qubit hyperentangled state in three degrees of freedom. If we
substitute the SPDC source in Ref.\cite{Hyper-fruquency} for the PDC
source in Ref.\cite{Simon}, we can produce the hyperentanglement
with the following form:
\begin{eqnarray}
\frac{1}{2\sqrt{2}}(|HH\rangle+|VV\rangle)\cdot(|\omega_{1}\omega_{2}\rangle
+
|\omega_{2}\omega_{1}\rangle)\cdot(|a_{1}b_{1}\rangle+|a_{2}b_{2}\rangle).\nonumber\\
\label{hyperentanglement}
\end{eqnarray}
Here, $H$ ($V$) represents the horizontal (vertical) photon
polarization, $\omega_{1}$ $(\omega_{2})$ represents the frequency
of the signal (idler) photon, and  $a_{1}b_{1}$ $(a_{2}b_{2})$
represents the spatial mode of photons.

In this article, we will present a deterministic EPP with
hyperentangled states in the form of Eq.(\ref{hyperentanglement}).
The two parties in quantum communication, say Alice and Bob, can get
a maximally entangled photon pair from each hyperentangled state in
this EPP, which is, in essence, different from all the existing
conventional EPPs \cite{Bennett1,Deutsch,Pan1,Simon,shengpra}. The
deterministic feature of our protocol will improve the efficiency of
long-distance quantum communication exponentially as the
conventional EPPs will consume entangled quantum resources
exponentially for obtaining some maximally entangled states. Also,
this EPP  can accomplish the complete nonlocal Bell-state analysis.

The article is organized as follows. In Sec.II A, we describe the
principle of the deterministic entanglement purification for
bit-flip errors with spatial entanglement. In Sec.II B, the
purification for phase-flip errors is discussed. In Sec.III, we
discuss the method of nonlocal Bell-state analysis with
hyperentanglement. In Sec.IV, we analyze the essence of entanglement
purification. In Sec.V, we discuss the present EPP in a practical
transmission. A discussion and a summary are given in Sec.VI.

\section{Deterministic entanglement purification}

It is well known that the most important application of EPPs is in
constructing quantum repeaters for long-distance quantum
communication in a noisy channel
\cite{repeater,DLCZ,sangouard1,simon1,zhao}. In order to connect the
adjacent nodes, the two parties in quantum communication should
first transmit their photons in a noisy channel and then connect
them with quantum entanglement swapping. Usually, the channel noise
 will degrade the entanglement of photon pairs.
Also, the imperfect operations will disturb the entanglement of
quantum systems. Now, let us start the explanation of our
entanglement purification scheme by discussing an ordinary example.
During a quantum-signal transmission, polarization degree of freedom
suffers from the channel noise as both the spatial degree of freedom
and the frequency degree of freedom are more stable than
polarization. The previous experiments showed that the polarization
entanglement is quite unsuitable for transmission over distances of
more than a few kilometers in an optical fiber \cite{rmp}. For
example, Naik \emph{et al.} demonstrated the Ekert protocol
\cite{Ekert} over only a few meters \cite{experiment1,rmp}. Also,
they observed the quantum bit error rate (QBER) increase to 33\% in
the experimental implementation of the six-state protocol
\cite{sixstate1,sixstate2}. For frequency coding
\cite{experiment2,experiment5,experiment3,experiment4,frequencystable1,frequencystable2},
for example, the Besancon group performed a key distribution over a
20-km single-mode optical-fiber spool. They recorded a QBER$_{opt}$
contribution of approximately 4\%, and estimated that 2\% could be
attributed to the transmission of the central frequency by the
Fabry-Perot cavity \cite{experiment5}. The experiment by
Min$\acute{a}\breve{r}$ \emph{et al.} \cite{experiment6} for
phase-noise measurements showed that in a realistic environment, the
phase in long fibers (several tens of km) remains stable, which is
an acceptable level for time on the order of 100$\mu s$. The phase
stabilization is relevant for the quantum repeaters in installed
optical fiber networks. In Simon's protocol \cite{Simon}, they also
performed an EPP using spatial entanglement to purify polarization
entanglement based on the good mode overlap on the PBS and phase
stability that were achieved in previous experiments.

In fact, that frequency and spatial entanglement absolutely  do not
suffer from the noise is only a hypothesis and  is unpractical. Here
we only use it to show the basic principle for our entanglement
purification process. We also will discuss the entanglement
purification under a realistic environment.  Moreover, other degrees
of freedom, such as time-bin, which is more robust than
polarization, can also be used to implement this scheme
\cite{experiment7,experiment8}. In the discussion section of this
article, we will show that the entanglement purification
 essentially performs entanglement transformation
between different degrees of freedom, that is,  transforms robust
entanglement in channel transmissions (frequency and spatial) into
 easily manipulatable entanglements (polarization).

Under the hypothesis mentioned previously, the  entanglement
purification in the present scheme is divided into two steps, that
is, purification for bit-flip errors and that for phase-flip errors.
We discuss them in detail in this section as follows.

\subsection{Deterministic purification for bit-flip errors}

After the transmission, the hyperentangled state of
Eq.(\ref{hyperentanglement}) will become a mixed one in
polarization:
\begin{eqnarray}
\rho_{p}&=& a|\Phi^{+}\rangle_{AB}\langle\Phi^{+}| + b|\Phi^{-}\rangle_{AB}\langle\Phi^{-}|\nonumber\\
&+& c|\Psi^{+}\rangle_{AB}\langle\Psi^{+}| +
d|\Psi^{-}\rangle_{AB}\langle\Psi^{-}|.\label{polarizationentanglement}
\end{eqnarray}
Here $a+b+c+d=1$ and  $\rho_{p}$ is the mixed part of
Eq.(\ref{hyperentanglement}) in polarization.
$|\Phi^{\pm}\rangle_{AB}$ and $|\Psi^{\pm}\rangle_{AB}$ are the
four Bell states for an entangled photon pair $AB$:
\begin{eqnarray}
\vert \Phi^{\pm} \rangle_{AB} =\frac{1}{\sqrt{2}}(\vert H\rangle_A
\vert H \rangle_B \pm \vert V\rangle_A\vert V\rangle_B),\\
\vert \Psi^{\pm} \rangle_{AB} =\frac{1}{\sqrt{2}}(\vert H\rangle_A
\vert V \rangle_B \pm \vert V\rangle_A\vert H\rangle_B).
\end{eqnarray}
After the transmission, the initial state becomes
\begin{eqnarray}
\rho=\rho_{p} \rho_{f} \rho_{s}\label{totalentanglement}
\end{eqnarray}
with one photon belonging to Alice and the other belonging to Bob.
Here
$\rho_{f}=\frac{1}{2}(|\omega_{1}\omega_{2}\rangle+|\omega_{2}\omega_{1}\rangle)(\langle\omega_{1}\omega_{2}|
 + \langle\omega_{2}\omega_{1}|)$ and
$\rho_{s}=\frac{1}{2}(|a_{1}b_{1}\rangle+|a_{2}b_{2}\rangle)(\langle
a_{1}b_{1}|+\langle a_{2}b_{2}|)$. We also let
$|\Psi_{f}\rangle=\frac{1}{\sqrt{2}}(|\omega_{1}\omega_{2}\rangle+|\omega_{2}\omega_{1}\rangle)$
and
$|\Phi_{s}\rangle=\frac{1}{\sqrt{2}}(|a_{1}b_{1}\rangle+|a_{2}b_{2}\rangle)$.

\begin{figure}[!h]
\begin{center}
\includegraphics[width=8cm,angle=0]{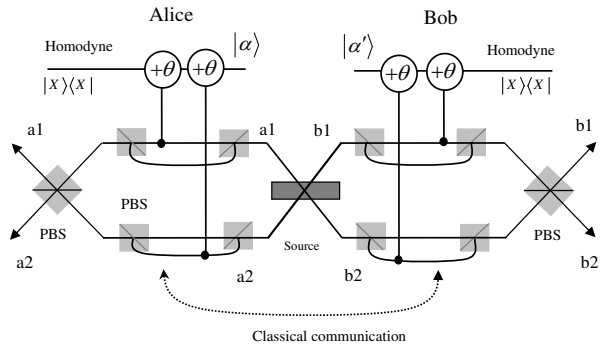}
\caption{Schematic drawing of the principle of bit-flip error
purification. The source emits the entangled pair with the form of
Eq.(1). One member of pair has been sent to Alice and the other to
Bob. Both Alice and Bob perform  X homodyne measurements on their
coherent beams $\vert \alpha\rangle$ and  $\vert \alpha'\rangle$,
respectively, and compare the results via classical communication.
If their results are different, they need to perform a bit-flipping
operation to correct this error. Otherwise, there are no bit-flip
errors. PBS: polarizing beam splitter.}
\end{center}
\end{figure}

There are admixtures of the unwanted states $|\Phi^{-}\rangle_{AB}$
and $|\Psi^{\pm}\rangle_{AB}$. We note that the state
$|\Phi^{+}\rangle_{AB}$ becoming $|\Psi^{+}\rangle_{AB}$ is a
bit-flip error, becoming $|\Phi^{-}\rangle_{AB}$ is a phase-flip
error, and both a bit-flip error and a phase-flip error take place
when $|\Phi^{+}\rangle_{AB}$ becomes $|\Psi^{-}\rangle_{AB}$. From
Eqs.(\ref{polarizationentanglement}) and (\ref{totalentanglement}),
the original state can be seen as a probabilistic mixture of four
pure states: with a probability of $a$ the photon pair in the state
$|\Phi^{+}\rangle |\Psi_{f}\rangle  |\Phi_{s}\rangle$, with a
probability of $b$ the pair in the state of $|\Phi^{-}\rangle
|\Psi_{f}\rangle  |\Phi_{s}\rangle$, with the probability of $c$ and
$d$ in $|\Psi^{\pm}\rangle |\Psi_{f}\rangle |\Phi_{s}\rangle$. The
whole task of purification is to correct the bit-flip and the
phase-flip errors. So this scheme includes two steps, one for
bit-flip error correction and the other for  phase-flip error
correction.

The principle of our scheme for  bit-flip error correction is shown
in Fig. 1, where $+\theta$ represents a cross-Kerr nonlinear medium
which will make the coherent state $\vert \alpha\rangle$  pick up a
phase shift $\theta$ when it and one photon couple with the medium
\cite{QND4,QND1,QND3}. We now  consider the  combinations
$|\Phi^{+}\rangle |\Psi_{f}\rangle |\Phi_{s}\rangle$ and
$|\Psi^{+}\rangle |\Psi_{f}\rangle |\Phi_{s}\rangle$. Let us first
discuss the state $|\Phi^{+}\rangle |\Psi_{f}\rangle
|\Phi_{s}\rangle$. In Fig. 1, the  items $|HH\rangle
(|\omega_{1}\omega_{2}\rangle+|\omega_{2}\omega_{1}\rangle)
|a_{1}b_{1}\rangle$ and $|VV\rangle (|\omega_{1}\omega_{2}\rangle +
|\omega_{2}\omega_{1}\rangle) |a_{2}b_{2}\rangle$ make the two
coherent beams $|\alpha\rangle$ and $|\alpha\rangle'$ obtain the
same phase shift of $\theta$, which can be detected by Alice and Bob
with an $X$ homodyne measurement \cite{QND4,QND1,QND3}. Finally,
coupled by the two polarizing beam splitters (PBSs), they will emit
from $a_{2}b_{2}$. The whole state becomes
$\frac{1}{2}(|HH\rangle+|VV\rangle)
(|\omega_{1}\omega_{2}\rangle+|\omega_{2}\omega_{1}\rangle)$.
Following by the same principle, we can also get the state
$\frac{1}{2}(|HH\rangle+|VV\rangle)
(|\omega_{1}\omega_{2}\rangle+|\omega_{2}\omega_{1}\rangle)$ from
$a_{1}b_{1}$  if both  Alice and Bob get no phase shifts. In the
case of $|\Psi^{+}\rangle |\Psi_{f}\rangle |\Phi_{s}\rangle$, it
never leads to the same phase-shift case. If Alice gets the phase
shift of $\theta$ and Bob gets no phase shift, it means that the
proceeding state is changed to $\frac{1}{2}(|HV\rangle+|VH\rangle)
(|\omega_{1}\omega_{2}\rangle+|\omega_{2}\omega_{1}\rangle)$. The
photon which belongs to Alice is in the mode of $a_{2}$, and which
belongs to Bob is in $b_{1}$. There is another case for Alice and
Bob. That is, Alice gets no phase shift and Bob gets $\theta$. The
corresponding state is also
$\frac{1}{2}(|HV\rangle+|VH\rangle)\cdot(|\omega_{1}\omega_{2}\rangle+|\omega_{2}\omega_{1}\rangle)$,
but with the spatial modes of $a_{1}$ and $b_{2}$.

By applying our purification procedure, Alice and Bob can easily
check the bit-flip error as they get different phase shifts with
their $X$ homodyne measurements on their coherent beams. The spatial
modes are also different, corresponding to the different collapsed
states , but which can be completely determined. Therefore, by
classical communication, if a bit-flip error occurs, Alice and Bob
will get rid of the bit-flip error by performing a bit-flip
operation $\sigma_{x}=|H\rangle\langle V|+|V\rangle\langle H|$. Next
we show our protocol works for the other cases. The case of
$|\Phi^{-}\rangle |\Psi_{f}\rangle |\Phi_{s}\rangle$ will get the
same result with $|\Phi^{+}\rangle |\Psi_{f}\rangle
|\Phi_{s}\rangle$. After the PBSs, the two photons will be either in
the upper modes $a_{1}$ and $b_1$ or in the lower modes $a_{2}$ and
$b_{2}$. For the case $|\Psi^{-}\rangle |\Psi_{f}\rangle
|\Phi_{s}\rangle$,  part of polarization both contains a bit-flip
error and a phase-flip error, so Alice and Bob will get the phase
shift of $\theta$ and 0 or 0 and $\theta$, respectively. It has an
analogy with the case of $|\Psi^{+}\rangle |\Psi_{f}\rangle
|\Phi_{s}\rangle$. We can also perform a bit-flip operation to
correct it. In this case, a phase-flip error  still remains.

\subsection{Deterministic purification for phase-flip errors}

So far, we have been talking about a bit-flip error for the mixed
state in polarization. By correcting this error, the initial state
in polarization becomes
\begin{eqnarray}
\rho_{p'}&=&(a+c)|\Phi^{+}\rangle_{AB}\langle\Phi^{+}|+(b+d)|\Phi^{-}\rangle_{AB}\langle\Phi^{-}|.
\end{eqnarray}
As we know, a phase-flip error cannot be purified directly, but it
can be transformed into a bit-flip error by Hadamard (H) operations.
In an optical system, it can be finished by a $\lambda/4$-wave plate
(QWP). By performing the H operations on the two photons with two
QWPs, Eq.(6) evolves:
\begin{eqnarray}
\rho_{p''}&=&(a+c)|\Phi^{+}\rangle_{AB}\langle\Phi^{+}|+(b+d)|\Psi^{+}\rangle_{AB}\langle\Psi^{+}|.
\label{polarizationstatestep2}
\end{eqnarray}
That is,  the initial state becomes
\begin{eqnarray}
\rho'=\rho_{p''} \rho_{f}. \label{statestep2}
\end{eqnarray}
It is interesting to find that the entanglement in frequency was not
affected during the  procedure discussed previously, but the spatial
entanglement is consumed for correcting bit-flip errors.

\begin{figure}[!h]
\begin{center}
\includegraphics[width=8cm,angle=0]{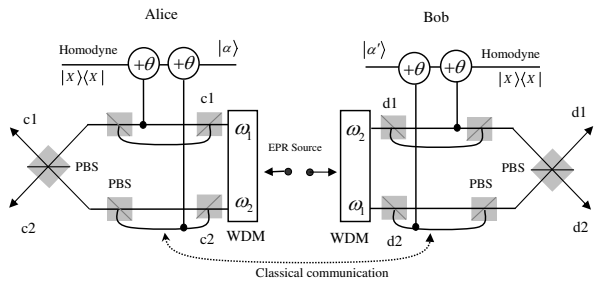}
\caption{Scheme showing the principle of phase-flip purification.
After two Hadamard operations, a phase-flip error is transformed
into a bit-flip error. Two WDMs are used to guide the the photons to
the different paths according to their frequencies. Similar to the
bit-flip error correction, if Alice and Bob get different phase
shifts, the phase-flip error occurs, and then they add a
bit-flipping operation to correct this error. Otherwise, there is no
phase-flip error. }
\end{center}
\end{figure}

Now we focus on the second step of our EPP:  correcting the
phase-flip error. In Fig. 2, two photons in the from of
Eq.(\ref{statestep2}) belong to Alice and Bob, respectively. Two
polarization-independent wavelength division multiplexers (WDMs) are
used to guide photons to different paths, according to their
frequencies. For example, in Alice's laboratory, it leads the photon
in $\omega_{1}$ to the mode $c_{1}$ and the photon in $\omega_{2}$
to the mode $c_{2}$. However, in Bob's laboratory, it leads the
photon in $\omega_{1}$ to the mode $d_{2}$ and the photon in
$\omega_{2}$ to the mode $d_{1}$.

From Eq.(\ref{polarizationstatestep2}), it follows that the original
state of the pairs can be seen as a probabilistic mixture of two
pure states: with a probability of $a+c$ the photon pair is in the
state $|\Phi^{+}\rangle |\Psi_{f}\rangle$ and with a probability of
$b+d$ the pair is in the state $|\Psi^{+}\rangle |\Psi_{f}\rangle$.
It is obvious that $|\Phi^{+}\rangle |\Psi_{f}\rangle$ leads to the
same phase shift for both $\theta$ and 0. They will be either in the
mode $c_{1}d_{1}$ with the state of
$\frac{1}{\sqrt{2}}(|H\omega_{2}\rangle_{A}|H\omega_{1}\rangle_{B}+|V\omega_{1}\rangle_{A}|V\omega_{2}\rangle_{B})$
or in the mode $c_{2}d_{2}$ with the state of
$\frac{1}{\sqrt{2}}(|H\omega_{1}\rangle_{A}|H\omega_{2}\rangle_{B}+|V\omega_{2}\rangle_{A}|V\omega_{1}\rangle_{B})$.
This is the maximally entangled state we need. However, for
$|\Psi^{+}\rangle |\Psi_{f}\rangle$, it never leads to the same
phase shift. Alice will get $\theta$ while Bob will get 0 with the
state
$\frac{1}{\sqrt{2}}(|H\omega_{1}\rangle_{A}|V\omega_{2}\rangle_{B}+|V\omega_{2}\rangle_{A}|H\omega_{1}\rangle_{B})$
in the mode $c_{2}d_{1}$ or Alice  gets 0 and  Bob gets $\theta$
with the state
$\frac{1}{\sqrt{2}}(|H\omega_{2}\rangle_{A}|V\omega_{1}\rangle_{B}+|V\omega_{1}\rangle_{A}|H\omega_{2}\rangle_{B})$
in the mode $c_{1}d_{2}$. We can perform a bit-flip operation to get
rid of the errors. Finally, we will get the entangled state
$\frac{1}{\sqrt{2}}(|H\omega_{2}\rangle_{A}|H\omega_{1}\rangle_{B}+|V\omega_{1}\rangle_{A}|V\omega_{2}\rangle_{B})$
or
$\frac{1}{\sqrt{2}}(|H\omega_{1}\rangle_{A}|H\omega_{2}\rangle_{B}+|V\omega_{2}\rangle_{A}|V\omega_{1}\rangle_{B})$
with a deterministic spatial mode. With quantum frequency
upconversion, we  can erase distinguishability for frequency
\cite{erase-frequency}.

\section{Nonlocal Bell-state analysis}

Now  let us discuss the relationship between this entanglement
purification protocol and a nonlocal Bell-state analysis (NBSA). A
universal conclusion is that a completely local Bell-state analysis
with linear optics is not possible and one can get an optimal
success probability of  only $\frac{3}{4}$
\cite{vaidman,bellmeasurement1,bellmeasurement2}. Several works have
shown that with additional degrees of freedom such as timing and
momentum it is possible to distinguish the four Bell states locally
\cite{hyper-analysis2,hyper-analysis3,hyper-analysis4}. However,
compared with local BSA, NBSA cannot be completed with the
collective operations and they can only resort to local operation
and classical  communication (LOCC). Here we will show that with
hyperentangled states and quantum nondemolition measurement (QND),
we can also perform a complete NBSA. The difference is that we have
to need another two degrees of freedom in NBSA, but only one is
needed in local BSA. The initial state is a hyperentangled state
with the form $|\Phi^{\pm}\rangle |\Psi_{f}\rangle |\Phi_{s}\rangle$
or $|\Psi^{\pm}\rangle |\Psi_{f}\rangle |\Phi_{s}\rangle$. The whole
protocol is the same as our EPP discussed previously. The first step
for NBSA is shown in Fig. 1. If Alice and Bob get the same phase
shift, that is, if both get 0 or $\theta$, they can decide that the
photon pair should be in one of the two states $|\Phi^{\pm}\rangle$.
Subsequently, they add a H operation on each photon and then make
the second check of the phase shifts (shown in Fig. 2). If the phase
shifts are still the same, the state should be $|\Phi^{+}\rangle$;
otherwise, it is  $|\Phi^{-}\rangle$. $|\Psi^{\pm}\rangle$ can be
distinguished in the same way. In the first step, if their phase
shifts are different, it must be one of the two states
$|\Psi^{\pm}\rangle$. Alice and Bob perform a bit-flip operation on
one of the photons and a H operation on each photon, which will
complete the transformation $|\Psi^{+}\rangle
|\Psi_{f}\rangle\longrightarrow |\Phi^{+}\rangle |\Psi_{f}\rangle$
and $|\Psi^{-}\rangle |\Psi_{f}\rangle\longrightarrow
|\Psi^{+}\rangle |\Psi_{f}\rangle$. In the second step, with the
help of frequency degree of freedom, if the outcomes of the
measurements on phase shifts are the
 same, Alice and Bob can conclude that the initial state should be
$|\Psi^{+}\rangle$; otherwise, it is  $|\Psi^{-}\rangle$.

\section{The essence of the present entanglement purification scheme}

In the previous works on entanglement purification, another
entanglement of degree of freedom, such as spatial entanglement, has
been used to purify the polarization entanglement of photon pairs
\cite{Simon,shengpra}. In Simon's protocol \cite{Simon}, the spatial
entanglement can  be used to purify a bit-flip error. After
consuming the resource of spatial entanglement, the phase-flip error
has to  be purified with the conventional method to repeat the same
procedure \cite{Bennett1,Deutsch,Pan1}. With another degree of
freedom of photons, we can accomplish a deterministic entanglement
purification.

Let us now discuss why our protocol can purify the mixed state
completely. From Eq.(\ref{polarizationentanglement}), we know that
there are two kinds of errors in the mixed state,  that is, one is a
bit-flip error and the other is a phase-flip error. The conventional
EPPs \cite{Bennett1,Deutsch,Pan1,Simon,shengpra} are used to purify
a bit-flip error. The phase-flip error cannot be purified directly,
but can be transformed into a bit-flip error. For the bit-flip
purification, Alice and Bob can check whether one of the pairs has a
bit-flip error. For instance, in Ref. \cite{Pan1}, the error
corresponds to the cross-combinations of $|\Phi^{+}\rangle_{AB}
|\Psi^{+}\rangle_{AB}$ and $|\Psi^{+}\rangle_{AB}
|\Phi^{+}\rangle_{AB}$. However, there always exists another
possibility that both of the states have bit-flip errors, which
corresponds to $|\Psi^{+}\rangle_{AB} |\Psi^{+}\rangle_{AB}$. In
this case, Alice and Bob cannot pick up these corrupt states, and
always keep them in the quantum systems that remain for a next
purification. That is, a bit-flip error cannot be purified
completely. Neither can a phase-flip error. They cannot make the
remaining ensemble  reach an indeed pure state. In Simon's protocol
\cite{Simon}, they revealed that another kind of entanglement can
also be used to purify the polarization entanglement state. In their
protocol, they can correct the bit-flip error completely as the
spatial entanglement state is a maximally entangled perfect pure
state and is not effected by the channel noise. Their protocol does
not lead to the case that each two-photon pair has bit-flip errors
after their purification. Following the same principle, we use the
frequency degree of freedom for a phase-flip error correction. Also,
we can use another kind of degree of freedom to complete this task
if it does not suffer from the channel noise.

LOCC cannot increase the entanglement of quantum systems. Therefore,
the process of entanglement purification is essentially the
transformation of entanglement. In the previous works
\cite{Bennett1,Deutsch,Pan1}, the transformation is between the same
kind of entanglement, that is, polarization entanglement. So they
need to consume the less-entangled pairs largely. The previous work
of Simon \cite{Simon} and this protocol show that entanglement can
be transformed between some different degrees of freedom. We let the
initial state be a hyperentangled state, and it owns three kinds of
degrees of freedom. During the purification step, we consume the
entanglement in frequency and spatial degrees of freedom to get a
pure polarization entangled state.  Thus, the whole entanglement
purification process does not need to consume the photon pairs but
to consume other degrees of freedom of entanglement.

\section{Entanglement purification in a practical transmission}

We have discussed our deterministic entanglement purification scheme
in the case where there are two degrees of freedom that are
insensitive to channel noise. We use the spatial freedom and the
frequency freedom of photons as an example to describe the principle
of our scheme. Of  course, the main experimental requirement of this
scheme is the phase stability if we use the spatial entanglement and
the frequency entanglement to purify the polarization entanglement.
This requirement may limit the distance of the quantum
communication. However, this scheme can be adapted to the case of
energy-time entanglement, which would allow the two parties
inquantum communication to be a father
apart\cite{Simon,longdistance}.

Now, let us discuss the present entanglement purification scheme
with a practical transmission based on the spatial entanglement and
the frequency entanglement.

In a practical transmission for long-distance quantum communication,
the relative phase between two different spatial modes is sensitive
to path-length instabilities, which may be caused by the fiber
length dispersion, or  atmospheric fluctuation in a free-space
transmission. In this way,  not only might  part of the polarization
of the hyperentangled state  become a mixed state shown in
Eq.(\ref{polarizationentanglement}), the entanglement in spatial
mode may become $\frac{1}{\sqrt{2}}(|a_1 b_1\rangle + e ^{i\Delta
\phi_{s}}|a_2  b_2\rangle)$ after transmission. The relative phase
between the different spatial modes is denoted by $\Delta
\phi_{s}=k\Delta x$. Here $k$ is the wave vector of the photons and
$\Delta x$ is the path-length dispersion between the two spatial
modes with $\Delta x=x_{a_1b_1}-x_{a_2b_2}$. That is to say, the
spatial entanglement will pick up a phase shift $\Delta \phi_{s}$.

Approximatively, the frequency entanglement has similar features to
the spatial entanglement. That is, it may become
$\frac{1}{\sqrt{2}}(|\omega_{1}\omega_{2}\rangle + e ^{i\Delta
\phi_{f}} |\omega_{2}\omega_{1}\rangle)$ after transmission. Here
$\Delta \phi_{f}$ is the phase dispersion coming from the different
frequencies.

After a practical transmission, the initial state may become
\begin{eqnarray}
\rho'=\rho_{p} \rho'_{f} \rho'_{s},\label{totalentanglementnoise}
\end{eqnarray}
where
\begin{eqnarray}
\rho'_{f} &=& \frac{1}{2}(|\omega_{1}\omega_{2}\rangle +
 e ^{i\Delta
\phi_{f}}|\omega_{2}\omega_{1}\rangle)(\langle\omega_{1}\omega_{2}|
 +  e ^{-i\Delta
\phi_{f}}\langle\omega_{2}\omega_{1}|) \nonumber 
\end{eqnarray}
and
\begin{eqnarray}
\rho'_{s} &=& \frac{1}{2}(|a_{1}b_{1}\rangle + e ^{i\Delta
\phi_{s}}|a_{2}b_{2}\rangle)(\langle a_{1}b_{1}| + e ^{-i\Delta
\phi_{s}}\langle a_{2}b_{2}|).\nonumber
\end{eqnarray}
With the first step for the purification of bit errors (the same as
the case where the entanglements in the spatial and the frequency
degrees of freedom do not suffer from the phase fluctuation, as
shown in Fig. 1), if both Alice and Bob get the phase shift $\theta$
on their coherent beams, the photon pair is in the state
$\frac{1}{2}(|HH\rangle+
 e ^{i\Delta
\phi_{s}}|VV\rangle) (|\omega_{1}\omega_{2}\rangle + e ^{i\Delta
\phi_{f}}|\omega_{2}\omega_{1}\rangle)$ with the probability of
$\frac{A}{2}$ and in the state $\frac{1}{2}(|HH\rangle -
 e ^{i\Delta
\phi_{s}}|VV\rangle) (|\omega_{1}\omega_{2}\rangle+ e ^{i\Delta
\phi_{f}}|\omega_{2}\omega_{1}\rangle)$ with the probability of
$\frac{B}{2}$, and they will emit from the spatial modes $a_2b_2$.
If both Alice and Bob get the phase shift $0$, the photon pair will
emit from the spatial modes $a_1b_1$ and   will be in the state
$\frac{1}{2}( e ^{i\Delta \phi_{s}}|HH\rangle+ |VV\rangle)
(|\omega_{1}\omega_{2}\rangle + e ^{i\Delta
\phi_{f}}|\omega_{2}\omega_{1}\rangle)$ with the probability of
$\frac{A}{2}$ and in the state $\frac{1}{2}( e ^{i\Delta
\phi_{s}}|HH\rangle - |VV\rangle) (|\omega_{1}\omega_{2}\rangle + e
^{i\Delta \phi_{f}}|\omega_{2}\omega_{1}\rangle)$ with the
probability of $\frac{B}{2}$. When Alice gets the phase shift
$\theta$ and Bob gets $0$, the photon pair will emit from $a_2b_1$
and  will be in the state $\frac{1}{2}(|HV\rangle+
 e ^{i\Delta
\phi_{s}}|VH\rangle) (|\omega_{1}\omega_{2}\rangle + e ^{i\Delta
\phi_{f}}|\omega_{2}\omega_{1}\rangle)$ with the probability of
$\frac{C}{2}$ and in the state $\frac{1}{2}(|HV\rangle -
 e ^{i\Delta
\phi_{s}}|VH\rangle)\cdot(|\omega_{1}\omega_{2}\rangle + e ^{i\Delta
\phi_{f}}|\omega_{2}\omega_{1}\rangle)$ with the probability of
$\frac{D}{2}$. When Alice gets the phase shift $0$ and Bob gets
$\theta$, the photon pair will emit from $a_1b_2$ and   will be in
the state $\frac{1}{2}(e ^{i\Delta \phi_{s}}|HV\rangle+
 |VH\rangle) (|\omega_{1}\omega_{2}\rangle + e ^{i\Delta
\phi_{f}}|\omega_{2}\omega_{1}\rangle)$ with the probability of
$\frac{C}{2}$ and in the state $\frac{1}{2}(e ^{i\Delta \phi_{s}}
|HV\rangle -
 |VH\rangle)\cdot(|\omega_{1}\omega_{2}\rangle + e ^{i\Delta
\phi_{f}}|\omega_{2}\omega_{1}\rangle)$ with the probability of
$\frac{D}{2}$. With some unitary operations, Alice and Bob can make
the state of their photon pair in the polarization degree of freedom
be
\begin{eqnarray}
\rho'_{p'}&=&(a+c)|\Phi'^{+}\rangle_{AB}\langle\Phi'^{+}|+(b+d)|\Phi'^{-}\rangle_{AB}\langle\Phi'^{-}|.\nonumber\\
\end{eqnarray}
Here
\begin{eqnarray}
|\Phi'^{+}\rangle_{AB}&=&\frac{1}{\sqrt{2}}(|HH\rangle +
 e ^{i\Delta
\phi_{s}}|VV\rangle),\\
|\Phi'^{-}\rangle_{AB}&=&\frac{1}{\sqrt{2}}(|HH\rangle -
 e ^{i\Delta
\phi_{s}}|VV\rangle).
\end{eqnarray}
Also, the state $\rho'_{p'}$ can be rewritten under the basis
$\{\vert \Phi^+\rangle, \vert \Phi^-\rangle, \vert
\Psi^+\rangle,\vert \Psi^-\rangle \}$ as
\begin{eqnarray}
\rho'_{p'}&=&\frac{1}{2}[1+(a+c-b-d)\text{cos}\Delta\phi_s]|\Phi^{+}\rangle_{AB}\langle\Phi^{+}|\nonumber\\
&+& \frac{1}{2}[1-(a+c-b-d)\text{cos}\Delta\phi_s]|\Phi^{-}\rangle_{AB}\langle\Phi^{-}|.\nonumber\\
\end{eqnarray}
That is, all the bit-flip errors in the photon pair are corrected
completely and there are only phase-flip errors in the quantum
system.

After the purification for bit-flip errors, Alice and Bob can
transfer phase-flip errors into bit-flip errors with unitary
operations again. That is, the state of the photon pair becomes
\begin{eqnarray}
\rho''_{p'}&=&\frac{1}{2}[1+(a+c-b-d)\text{cos}\Delta\phi_s]|\Phi^{+}\rangle_{AB}\langle\Phi^{+}|\nonumber\\
&+& \frac{1}{2}[1-(a+c-b-d)\text{cos}\Delta\phi_s]|\Psi^{+}\rangle_{AB}\langle\Psi^{+}|.\nonumber\\
\end{eqnarray}
With the setup shown in Fig. 2, if both Alice and Bob get the phase
shift $\theta$, the photon pair will emit from $c_2d_2$ and will be
in the state
$\frac{1}{\sqrt{2}}(|H\omega_{1}\rangle_{A}|H\omega_{2}\rangle_{B}+
e ^{i\Delta
\phi_{f}}|V\omega_{2}\rangle_{A}|V\omega_{1}\rangle_{B})$. If both
Alice and Bob get the phase shift $0$, the photon pair will emit
from $c_1d_1$ and will be in the state $\frac{1}{\sqrt{2}}(e
^{i\Delta \phi_{f}}|H\omega_{1}\rangle_{A}|H\omega_{2}\rangle_{B}+
|V\omega_{2}\rangle_{A}|V\omega_{1}\rangle_{B})$. When Alice gets
the phase shift $\theta$ and Bob gets $0$, the photon pair will emit
from $c_2d_1$ and will be in the state
$\frac{1}{\sqrt{2}}(|H\omega_{1}\rangle_{A}|V\omega_{2}\rangle_{B}+e
^{i\Delta \phi_{f}}
|V\omega_{2}\rangle_{A}|H\omega_{1}\rangle_{B})$. When Alice gets
the phase shift $0$ and Bob gets $\theta$, the photon pair will emit
from $c_1d_2$ and will be in the state $\frac{1}{\sqrt{2}}(e
^{i\Delta \phi_{f}}|H\omega_{1}\rangle_{A}|V\omega_{2}\rangle_{B} +
|V\omega_{2}\rangle_{A}|H\omega_{1}\rangle_{B})$. With quantum
frequency up-conversion, Alice and Bob  can erase distinguishability
for frequency \cite{erase-frequency} and they will get the entangled
state $\frac{1}{\sqrt{2}}(|H\rangle_{A}|H\rangle_{B} +e ^{i\Delta
\phi_{f}} |V\rangle_{A}|V\rangle_{B})$ with some unitary operations.

As the case discussed in Sec.II, Alice and Bob can correct the
bit-flip errors in their photon pair completely. In the step for
purification of phase-flip errors, the two different frequency modes
will introduce a relative phase shift $\Delta \phi_{f}$ in the Bell
state. In theory, $\Delta \phi_{f}$ does not change if the channel
lengths do not fluctuate with time $t$. That is, Alice can get
approximatively the maximally entangled state
$\frac{1}{\sqrt{2}}(|H\rangle_{A}|H\rangle_{B} +e ^{i\Delta
\phi_{f}} |V\rangle_{A}|V\rangle_{B})$. With a phase compensation,
Alice and Bob will get the standard Bell state
$\frac{1}{\sqrt{2}}(|H\rangle_{A}|H\rangle_{B} +
|V\rangle_{A}|V\rangle_{B})$.

\section{Discussion and Summary}

There are two approximative methods  used in our scheme in a
practical transmission. One is the assumption that the phase
dispersion in frequency degree of freedom $\Delta \phi_{f}$ is
independent of that in spatial modes $\Delta \phi_{s}$. The other is
the invariability of $\Delta \phi_{f}$ with time $t$. Certainly, the
case in a practical experiment is more complicated than that with
these two assumptions.

For the state
\begin{eqnarray}
\vert \Phi\rangle_{fs} &\equiv&
\frac{1}{2}(|\omega_{1}\omega_{2}\rangle +
|\omega_{2}\omega_{1}\rangle)\cdot(|a_{1}b_{1}\rangle+|a_{2}b_{2}\rangle)\nonumber\\
&=&\frac{1}{2}(|\omega_{1}\omega_{2}\rangle|a_{1}b_{1}\rangle
+|\omega_{2}\omega_{1}\rangle|a_{1}b_{1}\rangle\nonumber\\
& & +|\omega_{1}\omega_{2}\rangle|a_{2}b_{2}\rangle
+|\omega_{2}\omega_{1}\rangle|a_{2}b_{2}\rangle),\nonumber
\end{eqnarray}
each term will pick up a relative phase in a practical transmission.
That is, the state $\vert \Phi\rangle_{fs}$ will become
\begin{eqnarray}
\vert\Phi\rangle'_{fs}&=&\frac{1}{2}[e^{\frac{i}{v}(\omega_{1}L_{a_1}
+ \omega_{2}L_{b_1})}|\omega_{1}\omega_{2}\rangle|a_{1}b_{1}\rangle \nonumber\\
&& +\;\;  e^{\frac{i}{v}(\omega_{2}L_{a_1} +
\omega_{1}L_{b_1})}|\omega_{2}\omega_{1}\rangle|a_{1}b_{1}\rangle\nonumber\\
&& +\;\;  e^{\frac{i}{v}(\omega_{1}L_{a_2} +
\omega_{2}L_{b_2})}|\omega_{1}\omega_{2}\rangle|a_{2}b_{2}\rangle\nonumber\\
&& +\;\;  e^{\frac{i}{v}(\omega_{2}L_{a_2} +
\omega_{1}L_{b_2})}|\omega_{2}\omega_{1}\rangle|a_{2}b_{2}\rangle ].
\end{eqnarray}
Here $v$ is the velocity of photons in quantum channel. $L_{a_1}$,
$L_{a_2}$, $L_{b_1}$, and $L_{b_2}$ are the channel lengths in the
spatial modes $a_1$, $a_2$, $b_1$, and $b_2$, respectively. When
\begin{eqnarray}
\omega_{2}L_{a_2}+\omega_{1}L_{b_2} &\gg&
\omega_{2}L_{a_1}+\omega_{1}L_{b_1}+\omega_{1}(L_{a_2}-L_{a_1})
\nonumber\\
&& + \omega_{2}(L_{b_2}-L_{b_1})-\omega_{2}L_{a_2} -
\omega_{1}L_{b_2},\nonumber
\end{eqnarray}
$\vert\Phi\rangle'_{fs} $ can be rewritten as
\begin{eqnarray}
\vert\Phi\rangle'_{fs}&\approx
&\frac{1}{2}(e^{\frac{i}{v}(\omega_{1}L_{a_1} +
\omega_{2}L_{b_1})}\nonumber\\
&&  \times \{|\omega_{1}\omega_{2}\rangle +
e^{\frac{i}{v}[(\omega_{2}-\omega_{1})L_{a_1} +
(\omega_{1}-\omega_{2})L_{b_1}]}|\omega_{2}\omega_{1}\rangle\}\nonumber\\
&& \times \{|a_{1}b_{1}\rangle +
e^{\frac{i}{v}[\omega_{1}(L_{a_2}-L_{a_1})
+ \omega_{2}(L_{b_2}-L_{b_1})]} |a_{2}b_{2}\rangle \}).\nonumber\\
\end{eqnarray}
That is, when $2(\omega_{2}L_{a_2} + \omega_{1}L_{b_2}
)-\omega_{2}L_{a_1}-\omega_{1}L_{b_1}-\omega_{1}(L_{a_2}-L_{a_1})-\omega_{2}(L_{b_2}-L_{b_1})\gg
0$,
 the phase dispersion in frequency degree of freedom
$\Delta \phi_{f}$ can be regarded as being independent of that in
spatial modes $\Delta \phi_{s}$.

If $\Delta \phi_{f}(t)$ fluctuates with time $t$ in a small region,
Alice and Bob will get an ensemble in the state
$\rho_e=F_f|\Phi^+\rangle\langle\Phi^+| +
(1-F_f)|\Phi^-\rangle\langle\Phi^-|$ after a phase compensation $e
^{i\Delta \phi_{f}(0)}$, and they can, in this case, purify this
ensemble for getting some high-fidelity entangled states with
conventional EPPs \cite{Bennett1,Deutsch,Pan1,shengpra}. Here
$F_f=\frac{1}{2T}\int_0^T (1+\text{cos}(\Delta_f(t)))\text{d}t$,
$1-F_f=\frac{1}{2T}\int_0^T (1-\text{cos}(\Delta_f(t)))\text{d}t$,
and $\Delta_f(t))\equiv \Delta \phi_{f}(t)-\Delta \phi_{f}(0)$.
Different from the case  with a fixed phase dispersion, Alice and
Bob can only correct completely  the bit-flip errors in the photon
pair in this case. On the one hand, they will remove the phase-flip
errors in the polarization degree of freedom. On the other hand,
they will introduce the phase-flip errors in the frequency degree of
freedom. The latter comes from the phase dispersion between the two
frequencies of photons. If the latter is smaller than the former,
this scheme can be used to depress the ratio of phase-flip errors.

If $\Delta \phi_{f}(t)$ fluctuates acutely with time $t$,
$F_f\approx \frac{1}{2}$ and the ensemble maybe become a completely
mixed state and the two parties cannot distill maximally entangled
states, the same as the conventional EPPs in the case that the
initial fidelity is smaller than $\frac{1}{2}$. In fact, a quantum
channel fluctuating with time $t$ acutely is unsuitable for
entanglement purification as the Bell state required is mixed with
other Bell states uniformly. As the spatial and  frequency degrees
of freedom of photons are more stable than polarization, the
conventional EPPs will  not work if the present scheme does not
work.

In the  process of describing the principle of our entanglement
purification scheme, we exploit the cross-Kerr nonlinearity to
construct the QNDs. In fact, we should acknowledge that, on the one
hand, a clean cross-Kerr nonlinearity is quite a controversial
assumption with current technology. As pointed out in
Refs.\cite{shapiro,shapiro1,banacloche}, the single-photon Kerr
nonlinearity may do not help quantum computation, and a large phase
shift via a "giant Kerr effect" with single-photon wave packets is
impossible. On the other hand, here  a cross-Kerr nonlinearity is
only used to make a parity check for two photons and a strong Kerr
nonlinearity is not required. Meanwhile, other elements can also be
used to construct QNDs \cite{QND5,QND6,QND7} for this scheme.

In summary, we have presented a deterministic entanglement
purification scheme for purifying an arbitrary mixed state in
polarization  with present technology. The biggest advantage of this
scheme is that it works in a deterministic way in principle. That
is,  two parties can obtain a maximally entangled polarization state
from each hyperentangled state, which will improve the efficiency of
long-distance quantum communication exponentially. This protocol can
also be used to do the complete nonlocal Bell-state analysis, which
reveals that this deterministic entanglement purification scheme is
equal to a complete nonlocal Bell-state analysis. In a practical
transmission, this scheme can be used to correct all bit-flip errors
approximatively and depress the phase-flip errors in the
polarization degree of freedom, which will make it more efficient
than conventional EPPs. We believe that this scheme might be very
useful in the realization of long-distance quantum communication in
the future as entanglement purification is a very important element
in a quantum repeater and a quantum network.

\section*{ACKNOWLEDGEMENTS}
This work is supported by the National Natural Science Foundation of
China under Grant No. 10974020, the Foundation for the Author of
National Excellent Doctoral Dissertation of P. R. China under Grant
No. 200723, and the Beijing Natural Science Foundation under Grant
No. 1082008.

\end{document}